\documentclass[prl,twocolumn,showpacs,preprintnumbers,amsmath,amssymb, superscriptaddress]{revtex4}
\usepackage{graphicx}% Include figure files
\usepackage{dcolumn}% Align table columns on decimal point
\usepackage{bm}% bold math
\usepackage{fancyhdr}
\usepackage{float}
\usepackage{ifthen}
\usepackage{eurosym}
\usepackage{wrapfig}
\usepackage{xfrac}

\begin{document}
\preprint{Pr$_2$Pt$_3$Ge$_5$}
\title{Complete decoupling of magnetic order and superconductivity\\ in a conventional superconductor}

\ \author{D.~G.~Mazzone}
\altaffiliation{daniel.mazzone@psi.ch}
\ \affiliation{Laboratory for Neutron Scattering and Imaging, Paul Scherrer Institut, 5232 Villigen PSI, Switzerland}

\ \author{R.~Sibille}
\ \affiliation{Laboratory for Scientific Developments and Novel Materials, Paul Scherrer Institut, 5232 Villigen PSI, Switzerland}

 \author{M.~Bartkowiak}
\ \affiliation{Laboratory for Scientific Developments and Novel Materials, Paul Scherrer Institut, 5232 Villigen PSI, Switzerland}

\ \author{J.~L.~Gavilano}
\ \affiliation{Laboratory for Neutron Scattering and Imaging, Paul Scherrer Institut, 5232 Villigen PSI, Switzerland}

\author{C.~Wessler}
\ \affiliation{Laboratory for Scientific Developments and Novel Materials, Paul Scherrer Institut, 5232 Villigen PSI, Switzerland}

\ \author{M.~M\r{a}nsson}
\altaffiliation{Present address: KTH Royal Institute of Technology, Electrum 229, SE-16440 Kista, Stockholm, Sweden}
\ \affiliation{Laboratory for Neutron Scattering and Imaging, Paul Scherrer Institut, 5232 Villigen PSI, Switzerland}

\ \author{M.~Frontzek}
\altaffiliation{Present address: Oakridge National Laboratory, Tennessee, USA}
\ \affiliation{Laboratory for Neutron Scattering and Imaging, Paul Scherrer Institut, 5232 Villigen PSI, Switzerland}

\ \author{O.~Zaharko}
\ \affiliation{Laboratory for Neutron Scattering and Imaging, Paul Scherrer Institut, 5232 Villigen PSI, Switzerland}

\ \author{J.~Schefer}
\ \affiliation{Laboratory for Neutron Scattering and Imaging, Paul Scherrer Institut, 5232 Villigen PSI, Switzerland}

\ \author{M.~Kenzelmann}
\altaffiliation{michel.kenzelmann@psi.ch}
\ \affiliation{Laboratory for Scientific Developments and Novel Materials, Paul Scherrer Institut, 5232 Villigen PSI, Switzerland}

\date{\today}% It is always \today, today,
             %  but any date may be explicitly specified

\begin{abstract}
Superconductivity and magnetic order strongly compete in many conventional superconductors, at least partly because both tend to gap the Fermi surface. In magnetically-ordered conventional superconductors, the competition between these cooperative phenomena leads to anomalies at magnetic and superconducting phase boundaries.  Here we reveal that in Pr$_2$Pt$_3$Ge$_5$ superconducting and multiple magnetic order are intertwined within the same $HT$-phase space, but remain completely decoupled. Our thermal conductivity measurements provide evidence for normal electrons in the superconducting phase from which magnetic order emerges with negligible coupling to electron bands that contribute to superconductivity.
\end{abstract}

% PACS, the Physics and Astronomy Classification Scheme.

\maketitle

\bigskip

\section{I. INTRODUCTION}

The coexistence of magnetic order and superconductivity has been investigated ever since Ginzburg theoretically studied the possibility of ferromagnetic superconductors over half a century ago \cite{Ginzburg1956}. While strong magnetic spin fluctuations are present and play an important role in many unconventional superconductors \cite{steglich2012, Steward1984, Monthoux2007}, magnetic correlations compete strongly with conventional superconductivity. In clean metals even a 1$\%$ concentration of magnetic impurities can result in a complete loss of conventional superconductivity \cite{Matthias1958}.

Coexistence of magnetic order and conventional superconductivity has been observed in metallic materials featuring localized magnetic moments. In such systems, the magnetic ions are well isolated from the electrons of the conduction bands and the direct exchange between the magnetic ions and the electrons at the Fermi level is weak. This reduces scattering of Cooper pairs so that conventional superconductivity and magnetism can coexist \cite{Baltensperger1963, Matsubara1983}. Magnetic superconductivity was first found in $R$Mo$_6$S$_8$, $R$Mo$_6$Se$_8$, $R$Rh$_4$B$_4$ and $R$Ni$_2$B$_2$C (with $R$ being selected rare-earth elements) \cite{Fertig1977, Ishikawa1977, Eisaki1994, Cava19941, Cava19942, Gupta2006, Maple2000}. The competition between localized magnetism and superconductivity gives rise to exotic phenomena, such as reentrant superconductivity and anomalous upper critical fields \cite{Pfleiderer2009, Mueller2001, Maple1995}. Competition between magnetism and superconductivity arises from magnetostatic interactions and from competition of electron states at the Fermi surface, making it difficult to find materials where conventional superconductivity and magnetic order do not compete. In this article, we demonstrate that in Pr$_2$Pt$_3$Ge$_5$ the magnetic and superconducting properties are completely decoupled. We identify Pr$_2$Pt$_3$Ge$_5$ as a unique conventional superconductor where magnetic order and conventional superconductivity do not compete. It is thus a good model magnetic superconductor to selectively study interactions leading to magnetic order and conventional superconductivity in the same material. 

Pr$_2$Pt$_3$Ge$_5$ belongs to the family of compounds with the general formula $R_2M_3X_5$, where $R$ is a rare-earth element, $M$ a transition metal and $X$ a $s$-$p$ metal \cite{Sung2012, Nakajiama2008, Singh2001, Ramakrishnan2001, Skanthakumar1998, Noguchi1976}. The crystal structure is satisfactorily refined in the space group $Ibam$ with lattice constants $a$~=~10.13~\AA, $b$~=~11.86~\AA~and $c$~=~6.23~\AA~\cite{Sung2012}. At zero field, Pr$_2$Pt$_3$Ge$_5$ displays a transition at relatively high temperature, $T_c$~=~7.8~K, where a macroscopic fraction of the sample condensates into a superconducting phase. Transport measurements suggest that Pr$_2$Pt$_3$Ge$_5$ is a multiple gap conventional superconductor with a Curie-Weiss temperature of $T_{CW}$~$\approx$~-40 and -55~K for $\vec{H}||c$ and $\vec{H}||ab$, respectively. Additionally, two antiferromagnetic (AF) transitions at $T_{N_1}$~=~3.5~K and $T_{N_2}$~=~4.2~K are found inside the superconducting condensate \cite{Sung2012}.

\section{I. RESULTS AND DISCUSSION}

\subsection{A. Experimental details}

Single crystals were synthesized using a Pt-Ge mixture as self-flux \cite{Sung2012}. High purity elements (with atomic parts of 1~Pr, 4~Pt and 20~Ge) were placed in an alumina crucible and heated to 1130~$^{\circ}$C in an evacuated and sealed quartz tube. After 40 hours the liquid was cooled to 850~$^{\circ}$C at a rate of 3~$^{\circ}$C/h. The crystals were separated by means of a centrifuge using quartz wool as a filter. Typical single crystals showed RRR values of 5 and a consistent superconducting volume fraction $v_s\approx$~30\%, measured via magnetization measurements upon applying 1 mT along [001].

Electrical resistivity, magnetization and neutron diffraction were measured on the same single crystal of size 1.2~x~1.2~x~5.3~mm$^3$ and mass $m$~$\approx$~64~mg. The single crystal used for thermal conductivity measurements had a mass of $m$~$\approx$~12~mg with a shape of 0.7~x~0.9~x~3.35~mm$^3$. Electrical resistivity measurements were performed using a four-probe method in a Quantum Design PPMS. Magnetization data were recorded for magnetic fields $\mu_0H$~=~0~-~7~T with $\vec{H}||$[100], [010] and [001] using a Quantum Design MPMS-XL SQUID magnetometer. Simultaneous measurements of the resistivity and the magnetization were made in the MPMS magnetometer. Thermal conductivity results were obtained using the one-heater-two-thermometer steady-state method. The heat current was fixed along the crystallographic [001] direction, whereas the magnetic field could be rotated in the $ab$-plane. In order to exceed the superconducting critical field of indium solder joint that affixed the crystal to the thermal base, conductivity was measured at a minimal field of $\mu_0H$ = 50 mT at temperatures below 1 K. Single crystal neutron scattering was performed using the diffractometers DMC and TriCS as well as on the triple-axis spectrometer TASP at the Swiss Spallation Neutron Source (SINQ) at the Paul Scherrer Institut, Villigen, Switzerland. Diffraction experiments on DMC were carried out between $T$~=~1.8~-~10~K with a neutron wavelength $\lambda$~=~3.808~\AA. The investigations on TriCS were performed with $\lambda$~=~2.316~ and 1.175~\AA~in a temperature range $T$~=~1.7~-~10~K using either a 4-circle cryostat with a Joule-Thompson insert or a vertical 6~T magnet for fields applied along [001]. Diffraction studies on TASP were carried out with $\lambda$~=~2.36~\AA~between $T$~=~1.6~-~10~K and for fields $\mu_0H$~=~0~-~3~T with $\vec{H}||$[100]. The magnetic structures were refined using FULLPROF \cite{Fullprof}.

\subsection{B. Superconducting properties}

\begin{figure}[tbh]
\includegraphics[width=\linewidth]{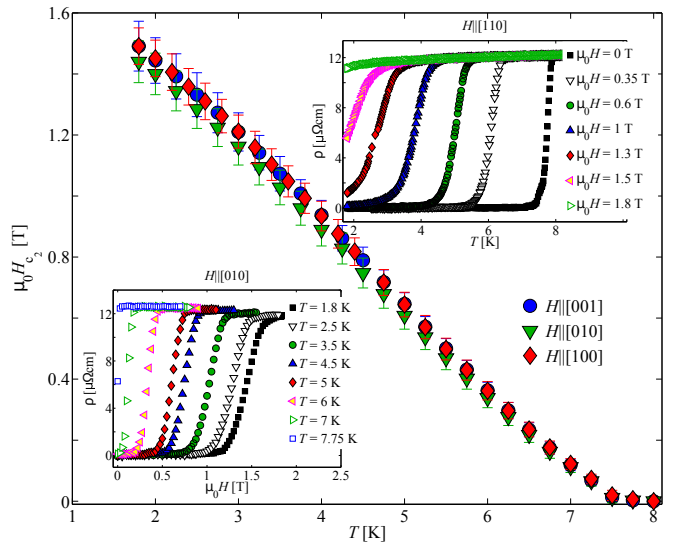}
\caption{Temperature dependence of the upper critical field, $H_{c_2}(T)$, measured via electrical resistivity. The insets show electrical resistivity data as a function of the temperature and magnetic field, respectively.}
\label{fig1}
\end{figure}

Fig.~\ref{fig1} displays the temperature dependence of the upper critical field, $H_{c_2}(T)$, for $\vec{H}||$[100], [010] and [001] obtained from field dependent resistivity measurements. At zero field a superconducting transition temperature $T_c$~=~7.8~K is found (see inset of Fig.~\ref{fig1}), in agreement with earlier reports \cite{Sung2012}. At low temperatures the upper critical field saturates at $\mu_0H_{c2}$(0)~$\approx$~1.6~T. The orbital limiting field of the superconducting condensate is estimated from the slope of the $\mu_0H_{c2}(T)$ near $T_c$ using $\mu_0H_{c2}^{orb}(0)$ = -0.73$|dH_{c_2}$/$dT|_{T_c}T_c$ = 1.5(1) T \cite{Helfand}.  This leads to a Ginzburg-Landau coherence length of $\xi_0$~$\approx$~148(5)~\AA~using $\mu_0H_{c2}$(0)~=~$\Phi_0$/2$\pi\xi_0^2$, with $\Phi_0$ the magnetic flux quantum \cite{Werthamer1966, Maki1966}. On contrary, the Pauli limiting field is calculated as $\mu_0H_{c2}^{P}(0)$ = $\Delta/\sqrt{g}$, where $g$ is the Lande factor and $\Delta$ the superconducting gap \cite{Maki1966}. Taking into account the reduced superconducting gap revealed by the heat capacity jump at $T_c$ ($\Delta C/\gamma T_c$ = 0.36 compared to the BCS value of 1.43 in isostructural La$_2$Pt$_3$Ge$_5$), the lower boundary of the Pauli limiting field in Pr$_2$Pt$_3$Ge$_5$ is approximated as $\mu_0H_{c2}^{P}(0)\approx$ 3.6 T \cite{Maki1966, Sung2012}. Thus, the system is characterized by an orbital limiting superconductor with an isotropic upper critical field, $H_{c_2}$($T$).

\begin{figure*}[tbh]
\includegraphics[width=\textwidth]{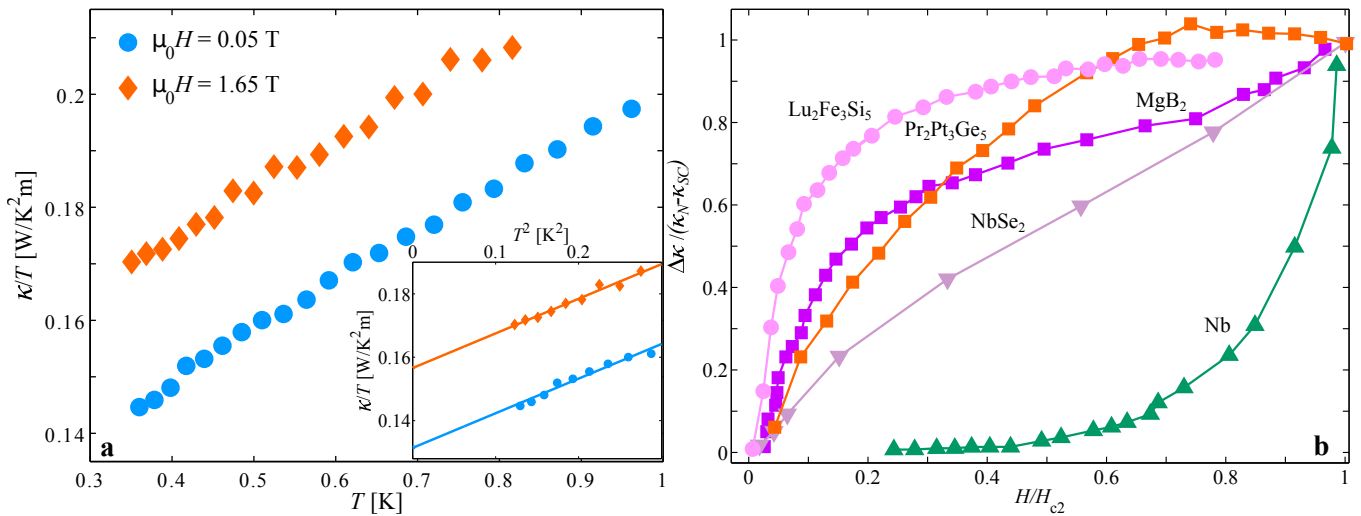}
\caption{\textbf{(a)} Temperature dependence of the thermal conductivity in the superconducting and normal state. \textbf{(b)} Field dependence of $\Delta\kappa/(\kappa_N-\kappa_{SC})$ =  ($\kappa(H)$ - $\kappa_{SC}$)/($\kappa_N$ - $\kappa_{SC})$ in Pr$_2$Pt$_3$Ge$_5$ and comparison with single and multiple gap superconductors.}
\label{heat1}
\end{figure*}

The low-temperature thermal conductivity divided by the temperature, $\kappa$($T$)/$T$, in the superconducting and normal state is shown in Fig. \ref{heat1}a. In the inset of Fig. \ref{heat1}a $\kappa/T$ is plotted against $T^2$. At very low temperatures the electron and phonon contribution of the thermal conductivity, $\kappa$~=~$\kappa_e$~+~$\kappa_{ph}$, are described by $\kappa$~=~$aT$~+~$bT^3$. The best fit to the data reveals $b$~=~0.109(8)~W/K$^4$m and $a$~=~0.1314(5)~W/K$^2$m at $\mu_0H$~=~50 mT. The non-vanishing electron contribution of the thermal conductivity at lowest temperatures suggests either nodal superconductivity or the presence of normal state electrons within the superconducting condensate. The second scenario can occur when not all electron bands develop a Fermi surface gap at the superconducting transition. Since Pr$_2$Pt$_3$Ge$_5$ is a superconductor with an isotropic upper critical field we attribute the non-vanishing residual thermal conductivity to conduction bands that do not contribute to superconductivity. The comparison of $\kappa_e$(50 mT) with $\kappa_e$(1.65 T) suggests that only 16.1(3)\% of the density of states is gapped below $T_c$. We point out that in the structurally related ternary iron silicades it has also been suggested based on specific heat measurements that electron bands remain normal in the superconducting state \cite{Vining1983}.

The field dependence provides further insight in the symmetry of the superconducting order parameter. We define $\Delta\kappa$ = $\kappa(H)$ - $\kappa_{SC}$ where $\kappa_{SC}$ denotes the thermal heat conductivity in the superconducting phase at minimal magnetic field, and $\kappa_N$ the thermal conductivity for $H$ = $H_{c_2}$. Fig. \ref{heat1}b shows $\Delta\kappa/(\kappa_N-\kappa_{SC})$ of Pr$_2$Pt$_3$Ge$_5$ measured for a fixed temperature, $T$ = 400 mK, as a function of magnetic field oriented in the $ab$-plane. This field dependence is compared with the multiple gap superconductors Lu$_2$Fe$_3$Si$_5$, MgB$_2$ and NbSe$_2$ as well as the $s$-wave superconductor Nb \cite{Lu, MgB2, NbSe2, Nb}. The field dependence of the thermal conductivity of Pr$_2$Pt$_3$Ge$_5$ is in dramatic contrast to the behavior of Nb, in which small fields hardly affect $\Delta\kappa/(\kappa_N-\kappa_{SC})$. A strong enhancement of $\Delta\kappa/(\kappa_N-\kappa_{SC})$ provides evidence for delocalized quasiparticles as found in multiple gap superconductors (see Fig. \ref{heat1}b). This comparison suggests that Pr$_2$Pt$_3$Ge$_5$ is a multigap superconductor and the change of slope suggests an upper critical field of $\mu_0H_{c_2}^s\approx$~0.3 T of the smaller gap. 

\subsection{B. Magnetic properties}

\begin{figure}[tbh]
\includegraphics[width=\linewidth]{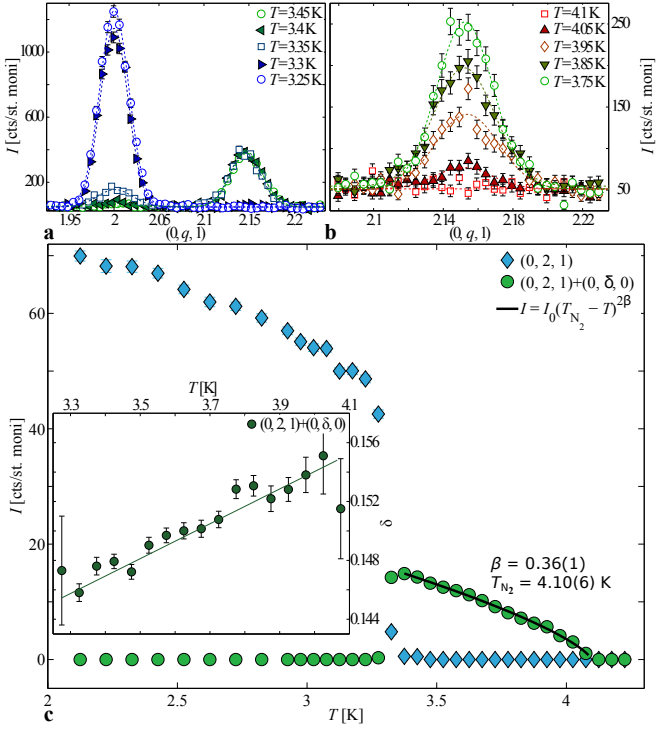}
\caption{$q$-scans of one CM and one ICM magnetic reflection along (0,~$q$,~1) for \textbf{(a)} 3.25~K~$\leq$~$T$~$\leq$~3.45~K and \textbf{(b)} 3.75~K~$\leq$~$T$~$\leq$~4.1~K. \textbf{(c)} Temperature dependent integrated intensities of (0,~2,~1) and (0,~2~+~$\delta$,~1). The ICM phase reveals a critical exponent $\beta$~=~0.36(1) and $T_{N_2}$~=~4.10(6) K. The inset depicts the incommensuration $\delta(T)$ as a function of temperature.}
\label{fig2}
\end{figure}

Using neutron diffraction we observed magnetic reflections below $T_{N_1}$ and $T_{N_2}$ at $\vec{Q}$~=~$\vec{G}$~$\pm$~$\vec{q}$ with $\vec{G}$ a  nuclear wave-vector. Representative data around the reciprocal lattice position (0,~2,~1) are shown in Fig.~\ref{fig2}a and b.  For $T_{N_1}$~$<$~$T$~$\leq$~$T_{N_2}$ an incommensurate (ICM) propagation vector $\vec{q_2}$~$=$~(0,~1~-~$\delta$,~0) with $\delta$~$\approx$~0.15 is found (see inset of Fig \ref{fig2}b). $\delta(T)$ decreases linearly with decreasing temperature for $T_{N_1}$~$<$~$T$~$<$~$T_{N_2}$ (see inset of Fig \ref{fig2}c). Below $T_{N_1}$ the magnetic structure abruptly adopts a commensurate (CM) wavevector, $\vec{q_1}$~$=$~(0,~1,~0), that breaks the body-centering of the unit cell. The propagation vectors of the CM and ICM structure are reminiscent of the magnetic phases of the isostructural compounds Tb$_2$Ni$_3$Si$_5$ and $R_2$Ni$_3$Si$_5$ ($R$~=~Tb, Pr and Nd) and Ce$_2$Ni$_3$Ge$_5$, respectively \cite{Skanthakumar1998, Durivault2002}.

Fig.~\ref{fig2}c shows the integrated intensities of the magnetic reflections (0,~2,~1) and (0,~2~+~$\delta$,~1) as a function of temperature. The gradual increase of magnetic intensity of the ICM magnetic structure reveals a second-order phase transition with a critical exponent $\beta$~=~0.36(1) and transition temperature $T_{N_2}$~=~4.10(6)~K (see black line in Fig.~\ref{fig2}c). Therefore, the data suggest a transition belonging to the universality class of either the 3D Heisenberg or the 3D model of XY spins. In contrast we observe a first-order type of transition  at $T_{N_1}$~=~3.4~K. We also find a hysteretic behavior with a small coexistence region of both magnetic structures, $\Delta T$~$\approx$~0.15~K (see Fig.~\ref{fig2}a and c) and $\Delta B$~$\approx$~60~mT at $T$~=~1.6~K and for $\vec{H}||$[100].

\begin{figure}[tbh]
\includegraphics[width=\linewidth]{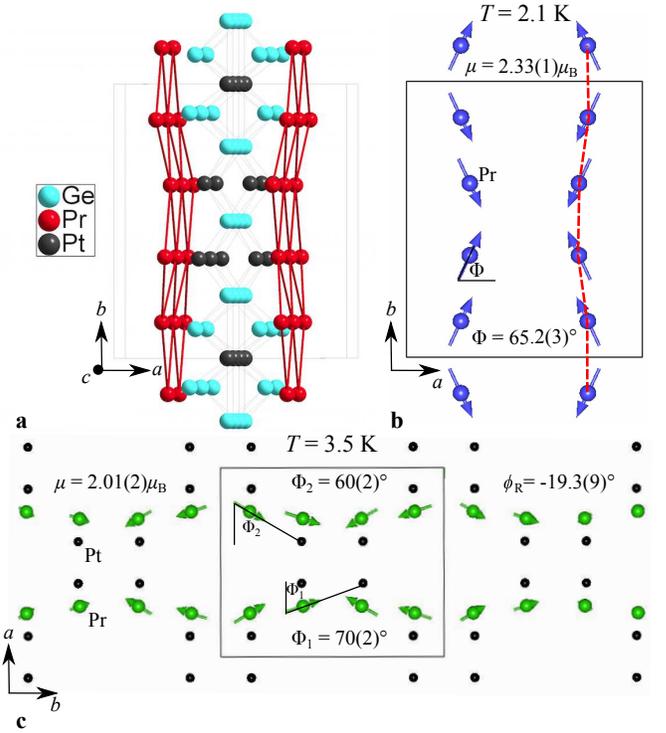}
\caption{\textbf{(a)} Crystal structure of Pr$_2$Pt$_3$Ge$_5$ with the distorted and buckled square lattice of Pr atoms in the $bc$-plane. Magnetic structure of \textbf{(b)} the CM and \textbf{(c)} the ICM phase in the $ab$-plane. The black box denotes the crystallographic unit cell. In the projection along $c$ two consecutive Pr atoms along the red dashed line are at different $c$-levels. The refinements of the CM and ICM structures reveal a non-collinear antiferromagnetic structure and an amplitude modulated non-collinear antiferromagnetic structure, respectively.} 
\label{fig3}
\end{figure}

The crystal structure of Pr$_2$Pt$_3$Ge$_5$ is displayed in Fig.~\ref{fig3}a. The nearest- and next-nearest-neighbors of the Pr atoms are found in the $bc$-plane. The mirror plane, orthogonal to the $c$-axis, restricts the magnetic moment either to the basal $ab$-plane or along the $c$-axis. The representational analysis for propagation vector $\vec{q_1}$ yields eight one dimensional irreducible representations with one magnetic orbit at site ($x$,~$y$,~$z$)~=~(0.26874,~0.36898,~0). Similarly, four one dimensional representations with two magnetic orbits, at sites ($x_1$,~$y_1$,~$z_1$)~=~(0.26874,~0.36898,~0) and ($x_2$,~$y_2$,~$z_2$)~=~(0.7313,~0.6310,~0) were found for propagation vector $\vec{q_2}$. The  best refinements of the magnetic Bragg peak intensities are obtained for the representations with the complex-valued basis vectors, $\vec{S}_j^{\vec{q_i}}$, shown in Table~\ref{table1} ($\vec{q_i}$ denotes the propagation vectors and  $j$ the positions of the magnetic atoms). The agreement factors, $R_F$, equal 4 and 3.5\% for the CM and the ICM phase, respectively.

\begin{table}
\centering
\begin{tabular}{l l l l}
\hline
\hline
&&&\\
\footnotesize{$\vec{S}_{(x,y,z)}^{\vec{q_1}}$}&\footnotesize{$\vec{S}_{(\textrm{-}x,\textrm{-}y,z)}^{\vec{q_1}}$}&\footnotesize{$\vec{S}_{(\textrm{-}x\textrm{+}\sfrac{1}{2},y\textrm{+}\sfrac{1}{2},\textrm{-}z)}^{\vec{q_1}}$}&\footnotesize{$\vec{S}_{(x\textrm{+}\sfrac{1}{2},\textrm{-}y\textrm{+}\sfrac{1}{2},\textrm{-}z)}^{\vec{q_1}}$}\\
\footnotesize{(M$_{x}$,~M$_{y}$,~0)}&\footnotesize{(-M$_{x}$,~-M$_{y}$,~0)}&\footnotesize{(M$_{x}$,~-M$_{y}$,~0)}&\footnotesize{(-M$_{x}$,~M$_{y}$,~0)}\\
&&&\\
\multicolumn{2}{l}{\footnotesize{$\vec{S}_{(x_{1,2},y_{1,2},z_{1,2})}^{\vec{q_2}}$}}&\multicolumn{2}{l}{\footnotesize{$\vec{S}_{(\textrm{-}x_{1,2}\textrm{+}\sfrac{1}{2},y_{1,2}\textrm{+}\sfrac{1}{2},\textrm{-}z_{1,2})}^{\vec{q_2}}$}}\\
\multicolumn{2}{l}{\footnotesize{(M$_{x_{1,2}}$,~M$_{y_{1,2}}$,~0)}}&\multicolumn{2}{l}{\footnotesize{(M$_{x_{1,2}}$,~-M$_{y_{1,2}}$,~0)$e^{2\pi i\textrm{0.074}}$}}\\
&&&\\
\hline
\hline
\end{tabular}
\caption{Complex-valued basis vectors, $\vec{S}_j^{\vec{q_i}}$, of the commensurate and the incommensurate structure for $\vec{q_1}$ = (0,~1,~0), $\vec{q_2}$ = (0,~0.852,~0), $j$ substitues ($x$,~$y$,~$z$) = (0.26874,~0.36898,~0), ($x_1$,~$y_1$,~$z_1$) = (0.26874,~0.36898,~0) and ($x_2$,~$y_2$,~$z_2$) = (0.7313,~0.6310,~0).}
\label{table1}
\end{table}

\begin{figure*}[tbh]
\includegraphics[width=\textwidth]{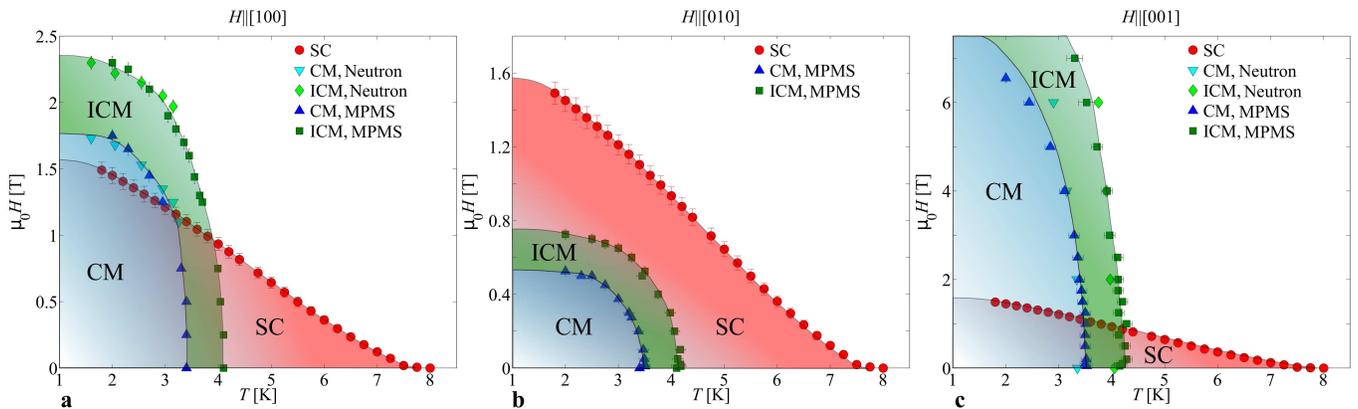}
\caption{$HT$-phase diagram of Pr$_2$Pt$_3$Ge$_5$ for \textbf{(a)} $\vec{H}||$[100], \textbf{(b)} $\vec{H}||$[010] and \textbf{(c)} $\vec{H}||$[001]. Red denotes the superconducting (SC), green the incommensurate (ICM) and blue the commensurate (CM) phase.}
\label{fig4}
\end{figure*}

The CM structure refinement at 2.1~K~=~$T$~$<$~$T_{N_1}$ reveals a non-collinear AF structure shown in Fig.~\ref{fig3}b. The structure displays an ordered magnetic moment of amplitude $\mu$~=~2.33(1)$\mu_B$. The magnetic moments are confined in the crystallographic $ab$-plane at an angle $\Phi$~=~65.2(3)$^{\circ}$ with respect to the $a$-axis. The total magnetic moment is reduced to a maximal amplitude of $\mu$~=~2.01(2)$\mu_B$ at $T$~=~3.5~K compared to $T$~=~2.1~K. The ICM magnetic structure at $T_{N_1}$~$<$~$T$~=~3.5~K~$<$~$T_{N_2}$ is depicted in Fig.~\ref{fig3}c. The refinement evidences an amplitude modulated structure with a local magnetic arrangement as in the CM phase. The two magnetic orbits are shifted by a relative angle $\phi_R$~=~-19.3(9)$^{\circ}$ and enclose the angles $\Phi_1$~=~70(2)$^{\circ}$ and $\Phi_2$~=~60(2)$^{\circ}$ with respect to the crystallographic $a$-axis.

Remarkably, the magnetic moments in the ICM phase point exactly along the direction of the Pt atoms in the basal $ab$-plane (see Fig.~\ref{fig3}c). This suggests that, for temperatures $T_{N_1}$~$<$~$T$~$<$~$T_{N_2}$, the magnetic structure is constrained by the crystal-field anisotropy. As a compromise between this anisotropy and the Rudermann-Kittel-Kasuya-Yosida (RKKY) exchange the magnetic moments orient along $\Phi$~=~($\Phi_1$~+~$\Phi_2$)/2 in the CM phase below $T_{N_1}$ \cite{Rudermann1954, Kasuya1956, Yosida1957}.  

\subsection{C. Interplay between superconductivity and magnetism}

In order to investigate the interplay between superconductivity and magnetism in 
Pr$_2$Pt$_3$Ge$_5$, we explored its $HT$-phase diagram (in Fig.~\ref{fig4}) using a combination of resistivity, magnetization and neutron diffraction measurements \cite{zusatz2}. While no anisotropy is found in the superconducting phase, a substantial anisotropy is detected for the magnetic phases. For $\vec{H}||[010]$ magnetic order is most strongly suppressed (see Fig.~\ref{fig4}b). This is in agreement with the magnetic refinement (c.f. Fig.~\ref{fig3}) and for magnetic moments with a large component along the $b$-axis, which will be strongly affected by magnetic fields along this direction. Here, both AF phases are fully embedded inside the superconducting phase and the absence of spin-flop transitions upon applying an external magnetic field indicates a strong single-ion anisotropy. In contrast, for $\vec{H}||[001]$ the AF phases survive to much higher fields compared to superconductivity (see Fig~\ref{fig4}c).  No noticeable changes of the magnetic structures are observed at $\mu_0H$~=~6~T for $\vec{H}||[001]$ and, on a relative error of 0.3\%, no anomalies in the intensities of the magnetic reflections are observed upon crossing the superconducting transition. For the hard axis, $\vec{H}||[001]$, the field generates a magnetization out of the basal $ab$-plane.

For $\vec{H}||$[100] the AF phases are partly inside and partly outside the superconducting phase (see Fig~\ref{fig4}a). The first-order transition from the CM to the ICM phase at low temperatures reaches a saturation at $\mu_0H$~$\approx$~1.73~T close to the superconducting phase at $\mu_0H_{c2}$(0)~$\approx$~1.6~T. However, a simultaneous measurement of the electrical resistivity and the magnetization rules out a collapse of both phases at the same field. 

\begin{figure}[tbh]
\includegraphics[width=\linewidth]{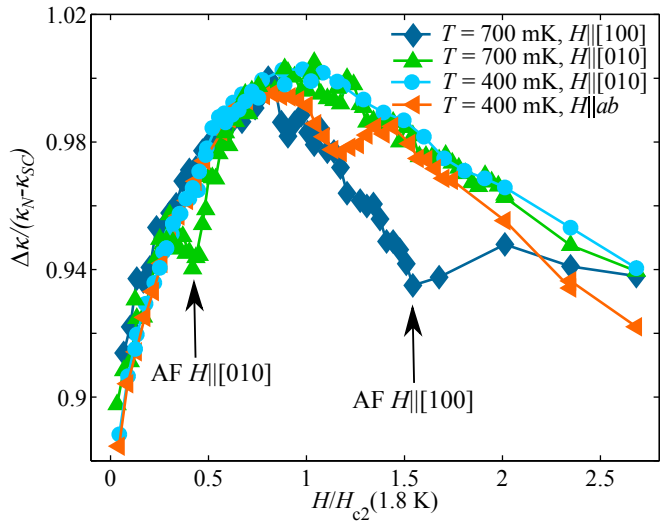}
\caption{Field dependence of the normalized electron contribution of the thermal conductivity for different temperatures and field directions.}
\label{heat2}
\end{figure}

Field and orientation dependent heat conductivity further identify the interplay between superconductivity and magnetism. Fig. \ref{heat2} displays $\Delta\kappa/(\kappa_N-\kappa_{SC})$ as a function of the normalized magnetic field, $H/H_{c_2}$, at $T$~=~400 and 700~mK and $\vec{H}||$[100], [010] and $ab$, respectively. Upon increasing the magnetic field the density of superconducting Cooper pairs is reduced, which causes an increase in $\Delta\kappa/(\kappa_N-\kappa_{SC})$ and results in a maximum at $H_{c_2}$. At the phase boundary between the superconducting and the ICM phase, $\Delta\kappa/(\kappa_N-\kappa_{SC})$ reveals a local minimum. This may be due to the change in rate of electron scattering by spin disorder at the AF phase boundary. At $T$~=~700~mK the reduction of the electron scattering rate is of the order of 6\% and less pronounced for lower temperatures (see Fig. \ref{heat2}). We speculate that this reduction is attributed to the reduction of thermal fluctuations at the phase transition for decreasing temperatures. The transition from the ICM to the CM structure is not seen in the thermal conductivity, suggesting that this transition involves only minimal changes that affect the quasi-particles at the Fermi surface. 

For $\vec{H}||[010]$, $\Delta\kappa/(\kappa_N-\kappa_{SC})$ is the same as for the other field directions, with the exception for fields close to $T_N$. This is strong evidence that the electrons that participate in the antiferromagnetic order are not involved with superconductivity. If the same electrons contributed to both superconductivity and antiferromagnetic order, one would expect an increase of thermal conductivity in the antiferromagnetically ordered phase with respect to the purely superconducting phase.

From the field dependent resistivity measurements a Ginzburg-Landau coherence length of $\xi_0$~$\approx$~148~\AA~is estimated using $\mu_0H_{c2}$(0)~=~$\Phi_0$/2$\pi\xi_0^2$, with $\Phi_0$ the magnetic flux quantum \cite{Werthamer1966, Maki1966}. Since the propagation vectors, $\vec{q_{1}}$ and $\vec{q_{2}}$, are much larger than $\xi_0^{-1}$, the average magnetic fields vanish on the size scale of a Cooper pair. This confirms that superconductivity is not expected to be suppressed by magnetostatic forces \cite{Matsubara21983}.  However, our data precludes any microscopic competition of electrons at the Fermi surface to contribute to superconductivity or magnetic order, because magnetic order has no influence on the bulk superconducting state and vice versa. 

This is in contrast to the well-studied magnetic superconductors $R$Mo$_6$S$_8$, $R$Mo$_6$Se$_8$, $R$Rh$_4$B$_4$ and $R$Ni$_2$B$_2$C \cite{Fertig1977, Ishikawa1977, Eisaki1994, Cava19941, Cava19942, Gupta2006}. In these materials the superconducting Cooper pairs are partly broken up by the RKKY-mediated magnetic states \cite{Pfleiderer2009, Mueller2001, Maple1995}. Most notable may be HoNi$_2$B$_2$C that features several magnetic structures inside the superconducting phase. As a consequence of the interaction between superconductivity and magnetism the $HT$-phase diagram exhibits a nearly suppressed superconducting phase at the onset of the ICM magnetic transitions \cite{Mueller2001}.

In Pr$_2$Pt$_3$Ge$_5$, there is no effect on the magnetic transition temperatures when the material becomes superconducting. Furthermore, thermal conductivity results reveal normal state electrons within the superconducting phase. This suggests that the Fermi surface that is involved in the formation of long-range order through the RKKY interaction does not contribute electrons into the superconducting condensate and remains ungapped below $T_c$. To our knowledge Pr$_2$Pt$_3$Ge$_5$ is the first conventional superconductor in which a competition between superconductivity and magnetism is completely absent. 

\section{III. SUMMARY}
 
In summary, thermal conductivity results provides evidence for a large density of normal state electrons within the superconducting phase of Pr$_2$Pt$_3$Ge$_5$ and suggests multigap superconductivity. Neutron diffraction data reveal a second-order phase transition at, $T_{N_2}$~$<$~$T_c$, into a magnetically-ordered phase inside the superconducting phase. For temperatures  $T_{N_1}$~$<$~$T$~$\leq$~$T_{N_2}$ and zero field an incommensurate, amplitude modulated non-collinear antiferromagnetic structure with $\vec{q_2}$~$\approx$~(0,~0.85,~0) is found, for which the magnetic moments are confined in the $ab$-plane with $\mu$(3.5~K)~$\approx$~2$\mu_B$. With decreasing temperature the magnetic structure transforms, at $T_{N_1}$,  into a commensurate, non-collinear antiferromagnetic structure with $\vec{q_1}$~=~(0,~1,~0) and $\mu$(2.1~K)~$\approx$~2.3$\mu_B$, without any detectable effect on superconductivity. The phase transition at $T_{N_1}$ is of a first-order with a similar moment configuration as in the incommensurate phase.  

As the key result of this study, we observe no significant anisotropy of $H_{c_2}(T)$ as a function of field direction, but a substantial anisotropy of the magnetic phases for different field directions. The interpenetrating $HT$-phase diagrams and field dependent thermal conductivity results along different crystal axes demonstrate that superconductivity and magnetism emerge from two separate, completely decoupled mechanisms. Thermal conductivity results suggest that different sheets of the Fermi surface are responsible for the superconducting pairing mechanism and for the RKKY interaction in Pr$_2$Pt$_3$Ge$_5$.

\section{ACKNOWLEDGMENTS}

We acknowledge the Paul Scherrer Institut for the allocated beam time on DMC, TriCS and TASP at SINQ. In addition, we thank the Swiss National Foundation (grant No. 200021\_147071 and 200021\_138018) and the European Community's Seventh Framework (FP7/2007-2013), grant No. 290605, COFUND: PSI-FELLOW. M. M. is partly supported by Marie Sk\l{}odowska Curie Action, International Career Grant through the European Union and the Swedish Research Council (VR), Grant No. INCA-2014-6426.

\bibliographystyle{apsrev}

\bibliography{Pr2Pt3Ge5}

\end{document}